\documentclass{elsart}
 
\usepackage{natbib,amssymb,amsmath,epsfig}

\begin{document}
\newcommand{\cal}{}
\newcommand{\apj}{ApJ}       
\newcommand{\apjs}{ApJS}     
\newcommand{\prd}{Phys. Rev. D}
\newcommand{\calu}{{\cal U}}
\newcommand{\calq}{{\cal Q}}
\newcommand{\bx}{{\rm \bf x}}
\newcommand{\bk}{{\bar{\kappa}}}
\begin{frontmatter}

\title{Gravitational Lensing of Epoch-of-Reionization Gas}
\author{Ue-Li Pen}
\address{Canadian Institute for Theoretical Astrophysics, University of
Toronto, M5S 3H8, Canada; pen@cita.utoronto.ca}
\date{\today}

\begin{abstract}
I present a weak lensing sensitivity estimate for upcoming high
redshift (epoch of reionization and beyond) 21cm surveys.  Instruments
such as PAST, LOFAR and SKA should be able to measure the weak lensing power
spectrum to precisions far exceeding conceivable optical surveys.
Three types of sources are detectable, which include the re-ionization
stromgren spheres, large scale structure, and minihalos.

Tomography allows the measurement of the time evolution of the dark
matter power spectrum.  Raw sensitivities allow measurement of many
cosmological parameters, including dark energy, neutrino mass and
cosmic equations of state, to percent accuracy.  It also has the
potential for inflationary gravity wave measurements.

Ultimate limits may be achievable through radio observations of $10^{18}$
minihalos.  Inflationary Hubble parameters $H_I$ down to $10^{-9} M_{\rm
Planck}$ can be detected through this effect.  Second order effects
may also be observable, allowing tests for backreaction and the
quantum mechanical origin of perturbations.

\end{abstract}


\end{frontmatter}

\section{Introduction}

Weak gravitational lensing has recently been established as a clean
tool to probe the distribution of all matter using only its
gravitational effects.  This can also be modelled from first
principles, and the matter power spectrum has already been measured to
some accuracy \citep{2003MNRAS.346..994P}.

Current measurements of the cosmic microwave background have resulted
in power spctra that are accurate at the percent level when
appropriately binned.  This constrains some combinations of
cosmological parameters to similar percent level accuracies.  But
there are many degeneracies in standard cosmological parameters
arising from purely CMB data.  Using the WMAP data alone, even
unpopular regimes are allowed, such as a closed model with low Hubble
constant\citep{2003ApJS..148..175S}.  The authors dismiss such models
as ``unreasonable'', primarily due to the non-standard required
value of $H_0=35$.  Nevertheless, one can find references in the
recent literature which predicted such low Hubble
constants\citep{1995Sci...267..980B}.

Some parameters, such as the absolute baryon density $\Omega_bh^2$ are
very well determined by the CMB alone, while many other physical
parameters require external data to constrain accurately, and thus have much
larger uncertainties.  Some such examples include the matter density
of the univere, equation of state, or neutrino mass.  Precision low
redshift measurements are also needed to break these degeneracies.
Galaxy distributions have been used successfully as a calibrator
\citep{2003ApJS..148..175S}, but its precision may have fundamental
limitations.  The external calibrators tend to be limited by
systematic errors in the models, whose error bars are difficult to
quantify.

Robust low redshift degeneracy breaking measurements have already come
from weak gravitational lensing \citep{2003PhRvL..90v1303C}.  Future
lensing surveys will signficantly improve on the statistical accuracy,
which is currently at the 10\% level.  Accuracies greater than 1\% are
in progress \citep{2003astro.ph..5089V}.  Much of the most interesting
information, such as the equation of state of the universe or neutrino
mass \citep{1999A&A...348...31C,2000PhRvL..84.1082F}, requires even
more precise measurements.  Multiple redshift measurements would be
invaluable.  Large physical scales are important to measure the linear
power spectrum, which are not affected by baryon feedback.  All of
these are very challenging for the current weak lensing strategy.  The
surface densities of galaxies on the sky is limited, and large sky
coverage is expensive.

The ideal source for weak lensing maps should be at high redshift, and
have structure on small scales.  The CMB satisfies the first
criterion, but not the second.  Faint galaxies partially satisfy both.
The epoch of reionization universe is a natural candidate to use as a
source screen to measure gravitational lensing.  It is far away, emits
brightly in the hydrogen hyperfine transition line, has structures on
many scales ranging from several arcminutes to less than a
milliarcsecond, a high surface density, and redshift information on
each source.  Several experiments are under construction, including
PAST (http://astrophysics.phys.cmu.edu/\~\ jbp/past6.pdf), LOFAR
\citep{2000SPIE.4015..328K} and SKA is in the design phase
\citep{2000pras.conf.....V}.  These instruments will measure this
source population to high accuracy.  Several more experiments,
including T-REX (http://orion.physics.utoronto.ca
/sasa/Download/poster/ casca\_poster.pdf) and CATWALK
(ftp://ftp.astro.unm.edu/pub/users/john/AONov03.ppt), may result in
earlier detections of re-ionization effects, but at lower
signal-to-noise.  The Canadian Large Adaptive Reflector CLAR
\citep{2000SPIE.4015...33C} will also measure this redshift range at
lower spatial resolution.  In this paper I present sensitivity
estimates for these experiments, and continue to estimate the
cosmological limits if one exhausted the information.

Some of these instruments do not have finalized designs and the cosmic
reionization history is not well known.  Therefore we only make rough
estimates, which we expect are probably good to a factor of two.

\section{Pre-reionization Weak Lensing}

We first estimate the strength of gravitational lensing for sources
at redshift $z_s=9$. 
We use the WMAP cosmological parameters \citep{2003ApJS..148..175S} with
$\Omega_0=0.27,\ \Omega_\Lambda=0.73,\ \sigma_8=0.84$.  The angular diameter 
distance to redshift 9 is $2.23c/H_0=6.7h^{-1}$ comoving Gpc.  One
arcsecond is $33 h^{-1}$ kpc (comoving).
\begin{figure}
\centerline{\epsfig{file=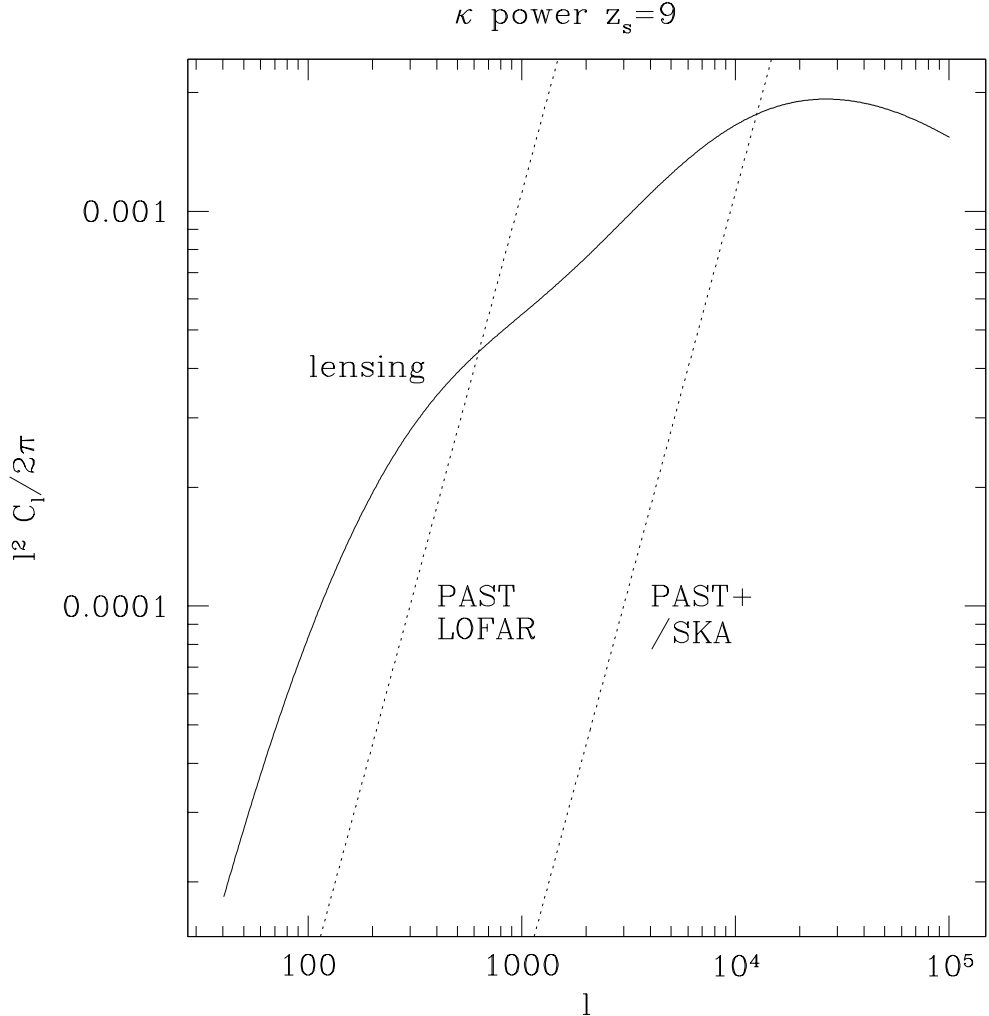, width=\columnwidth}}
\caption{The lensing power spectrum for sources at $z=9$.  The solid
line is the shear and convergence power spectrum.  
Images in the source plane are distorted by the square root of the power,
which is several percent. The left dotted line is the noise level
expected from PAST/LOFAR.  The dotted line on the right is the
noise expected from the second generation PAST+ and the SKA.
}
\label{fig:kps}
\end{figure}

Figure \ref{fig:kps} shows the lensing convergence ($\kappa$) power
spectrum computed by integrating the Limber equation over the
\citet{1996MNRAS.280L..19P} nonlinear matter power spectrum.  The
horizontal axis represents the angular scale, with $\theta \sim
2\pi/l$.  The vertical scale is the variance in the $\kappa$ power,
which corresponds to the magnification factor of point sources in the
weak lensing limit.  The plot also shows the Poisson noise from
PAST/LOFAR and the second generation PAST+/SKA.   

The most prominent feature is
the peak at $l\sim 10^4$.  This corresponds to an angular scale of
two arc minutes.  Current lensing surveys have signal to noise less
than unity at the peak of the spectrum, which prevents the
construction of direct dark matter maps.  At these higher redshifts,
the lensing power is a few parts in a thousand.  The rms amplification
or shear is several percent, which makes its measurement more
tractable than the one percent level fluctuations in current cosmic
shear surveys.  The variations are not linear to the statistical
accuracy, and strong lensing corrections can be considered.  These
have been presented, both in perturbative and non-linear form, by
\citet{2000ApJ...534L..19P}.

Before reionization, most of the baryonic matter in the universe is in
the form of neutral hydrogen.  After $z\lesssim 30$, the majority of
that gas gravitationally collapses with the dark matter into
virialized 'mini-halos'.  The cosmic microwave background has a
temperature $T_{\rm CMB}=2.73(1+z)$.  The gas is collisionally heated
to the virial equilibrium temperature of the minihalos, which is
between $10^2-10^4$K, so typically hotter then the CMB.  The average
density of a minihalo is 200 times the mean density.  At these
temperatures and densities, the spin temperature is tightly coupled to
the kinetic temperature, and decoupled from the CMB
\citep{1997ApJ...475..429M,2002ApJ...572L.123I,2003MNRAS.341...81I}.
In the limit of a spin temperature much higher than the CMB
temperature, the average surface brightness at mean density in
spontaneous emission on the 21cm line is $T_{21cm}=23$mK
\citep{1997ApJ...475..429M}.  These minihalos provide a fluctuating
source screen with a characteristic scale that can be used to measure
the gravitational lensing effect of the intervening dark matter.  In
the subsequent sections we will discuss the observability of this
effect.

\section{Measuring the power spectrum}

\subsection{High S/N}

Two regimes exist for measuring the weak lensing shear at high
redshift.  With high angular resolution and signal-to-noise, one
expects to resolve individual minihalos.  The characteristic mass is
$10^5 M_\odot$, and one expects $\sim 10^{18}$ such minihalos on the
sky.  Each virialized halo is approximately round, which can measure
the local shear with a signal-to-noise of order unity.  The
characteristic size of a halo is 1kpc, corresponding to 30 mas.  This
requires a baseline of $D=$13000km, which is resolvable with earth
sized baselines.  The corresponding spherical harmonic number is
$l\sim 10^7$. The temperature of such a halo is $\sim 10^3$K, with a
velocity width of about 3km/sec.  Virialized halos are about a factor
of 200 overdense in 3-D.  Along the line of sight the radial velocity
dispersion is comparable to the size of the infall region, so one only
expects a two dimensional density enhancement, which would be $\sim
40$.  The brightness temperature on a halo would then be $\sim 0.1$K.

An minimal observing strategy would aim to achieve a signal-to-noise
of 1 on the typical minihalo.  A lower signal to noise cannot identify
individual halos.  Higher signal-to-noise is expensive to achieve.
The actual accuracy on the lensing map reconstruction is boosted by
the number of halos along the same line of sight that one averages
over.  We use a simple model to determine signal-to-noise.  A filled
aperture telescope of diameter $D$ corresponding to the maximal
baseline is used as reference.  If the elements are distributed
uniformly within this diameter, we can treat the actual sensitivity as
just a dilution corresponding to a telescope with a very small
aperture efficiency $\eta_A=4A_{\rm eff}/(\pi D^2)$ where $A_{\rm
eff}$ is the collective effective area of all the elements.  The noise
in the map is then $T_{\rm sys}/(\eta_A \sqrt{t\Delta \nu})$, where we
take the integration time $t$ to be one year, and the bandwidth the
thermal width of each minihalo of 3km/sec, $\Delta \nu\sim 1.4$khz.
To obtain a signal-to-noise of unity on each halo in a year at a
system temperature of 200K, an effective aperture of 200,000 km$^2$ is
required.  As an interferometer, 200 elements with 40 km apertures
would suffice.  A half wave dipole has an effective area of $A_{\rm
eff} = 3\lambda^2/(8\pi)\sim 0.5 {\rm m}^2$, so 400,000km of wire are
needed to achieve this aperture.  Copper has a resistivity of
$1.7\times 10^{-8} \Omega \cdot m$, so a diameter of 0.1mm results in
a resistance much less than the impedance.  The wire requires 40 tons
of copper.  At the date of writing, one pound of copper costs US\$1.1,
which is a neglible cost factor.  The amplifiers, interconnects and
processing would dominate, but this cost reduces by Moore's law.

\subsection{Low S/N}

Initial experiments such as PAST and LOFAR will measure the
distribution of gas on linear scales, where the only observable are
Gaussian fluctuations, and reionization Stromgren spheres.  The
simplest way to measure the fluctuations is to construct a three
dimensional map, with the redshift as the third dimension.  On each
point of the angular map, we compute the variance of the density field
along the z dimension.  One way of achieving that is to Fourier
transform the spectrum, and measure the power in each Fourier mode.
In practice, some modelling would be needed to achieve appropriate
noise weights.  There are different ways of reconstructing a lensing
map from a 'fuzzy' noisy image.  If only one two dimensional image
were known, for example in the CMB, this can only be done if we have
knowledge of the statistics of the source screen.  The CMB, by being
Gaussian, allows such reconstruction using the connected four point
function which is zero for a stationary Gaussian process, but is
induced by lensing\citep{2003PhRvD..67h3002O}.  If the unlensed
properties are not known, it it presumably very difficult to
disentangle lensing from the intrinsic
statistics, but also see \citet{2004NewA....9..173C}.

Our reconstruction will use the intrinsic three dimensional
information that is available from redshifted 21cm structures.  This
gains us two kinds of observables: the intrinsic correlations of
matter are presumably statistically isotropic.  Lensing will change
that anisotropically.   Sources that are distant along the line
of sight but close in projection pass through the same gravitational
lensing screen, and will have induced cross correlations.  It is this
latter effect that we exploit.  We observe the structures at a
constant angular scale.  Weak lensing changes the mapping of angular
to physical scale slightly.  Properties of the source generically
depend on the physical scales at which they are measured, and thus
their statistics will also experience a slight change.  This is true
for all objects along the same line of sight, even the distant ones.
The variance in the beam, for example, will generally increase on
small scales.  This occurs in a correlated fashion for objects over a
range in source redshift.  The procedure now is to take a point in the
map, smooth the data along the redshift space axis, square that, and
compute the mean variance.  Now we can ask what the optimal smoothing
window would be.  The smoothing procedure is a convolution, which is
the same as Fourier transforming the line, squaring each Fourier mode,
and summing them weighted by a window.  Each of the Fourier modes is
uncorrelated because of stationarity (independent of Gaussianity), so
we do not need to worry about covariances between fourier modes.  We
presumably want the window which maximizes sensitivity to gravitational
lensing in the presence of noise.  If the data is noise dominated, we
weight each point by the ratio of signal to noise variance.  One 
needs to know the expected variance as a function of redshift, which
one can measure from the angular maps.

\begin{figure}
\centerline{\epsfig{file=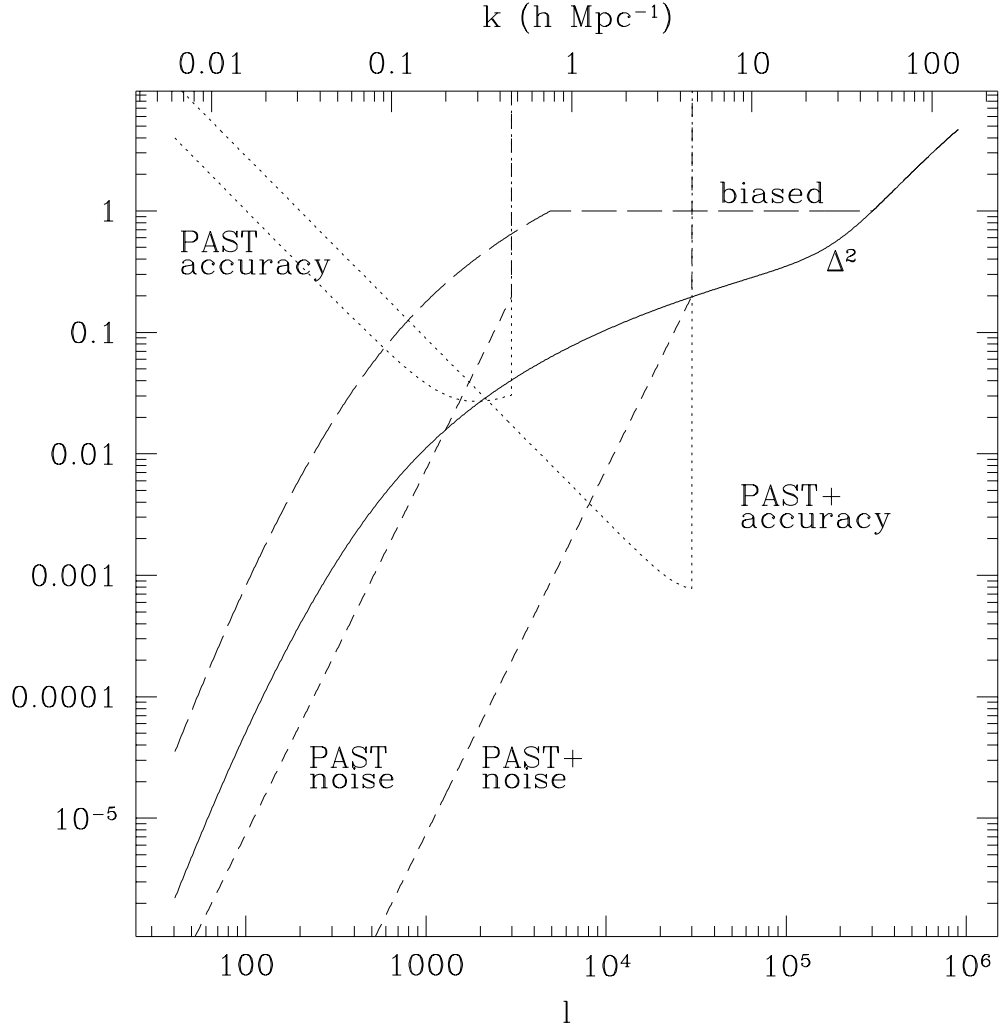, width=\columnwidth}}
\caption{Matter power spectrum measurement. 
Solid line: the matter power spectrum at $z=9$.  One expects the 21cm
emission to trace the total mass.  Dot-dashed line: power spectrum of
patchy re-ionization with a bias $b=4$.
Dashed lines: noise level expected
for PAST and second generation PAST+.  Dotted line: fractional
accuracy of power spectrum measurement
per logarithmic $l$ bin. }
\label{fig:mpower}
\end{figure}

This results in a two dimensional map $\sigma^2(\theta_x,\theta_y)$
whose entries are projected variances in the density of neutral
hydrogen.  We denote the three dimensional matter power spectrum by
$\Delta^2(k) \equiv k^3 P(k)/(4\pi)$.  It is shown in figure
\ref{fig:mpower}.  A pixel size $\theta$ maps to a spherical harmonic
number $l\sim2\pi/\theta$.  One can think of the variance as a measure
of $\Delta^2(l)$ shown in figure \ref{fig:mpower}.  Patchy
re-ionization can significantly boost the 21cm power by modulating the
neutral fraction.  Figure \ref{fig:mpower} shows a model of the
boosted power, using a $b=4$ bias model on large scales
\citep{2003ApJ...598..756S}.  Since the neutral fraction cannot be
modulated by more than unity, we limited the biased power to the
greater of unity and the actual matter power.

Gravitational
lensing has two effects on the variance.  It changes the area of each pixel
in the source plane.  Each pixel is a fixed solid angle on the sky.
The variance measures the power at the pixel scale.  As the physical size
corresponding to the pixel is increased or decreased
by gravitational lensing, the
same pixel probes a larger or smaller angular scale.  From Figure
\ref{fig:mpower} we see that moving to larger scales (smaller $l$) decreases
the variance in each pixel.

This variance map is related to the strength of lensing convergence.
The fractional change in the variance by a fractional change in
area is half the logarithmic derivative of the power spectrum:
\begin{equation}
\frac{\sigma^2({\bf x})}{\langle\sigma^2\rangle}-1
=\frac{\kappa({\bf x})}{2}
\frac{\partial \log \Delta^2(k)}{\partial \log k}.
\label{eqn:kappa}
\end{equation}
We expect to evaluate the derivative at the physical scale $k$ corresponding
to angular scale $l$ of the pixel on the sky.  From Figure \ref{fig:kps}
we see that the logarithmic derivative is of order unity.

The other observable from gravitational lensing is the gravitational
shear.  If one takes a collection of pixels on the sky, one can
construct a three dimensional two point correlation function.
Projecting the correlation function along the radial direction, we can
examine the anisotropy of the two point correlation function.  In the
absence of lensing, the 2-D two-point correlation function is
statistically rotation invariant on the sky.  Isovariance lines of the
correlation function are circles.  Gravitational shearing will distort
the correlation function on each patch to appear anisotropic.  The
ellipticity is a measure of the gravitational reduced shear. The
signal-to-noise of the two methods is comparable.  The benefit of the
shear procedure is that no prior knowledge of the power spectrum is
required.

\subsection{LOFAR/PAST}

We examine a rough sensitivity estimate for PAST and an idealized
LOFAR configuration.  For this purpose, we measure the gas power
spectrum at $z\sim 9$.  We assume a system temperature $T_{sys}=400$K
which is dominated by the galaxy, and a baseline of 2km as is planned
for PAST.  The 21cm line becomes 2.1m, and the telescope has an
effective resolution of 1/1000 radians if it were a single dish.  We
can approximate this to be a spherical harmonic number of $3000$.  The
angular scale corresponds to three arcminutes, and 7 $h^{-1}$ Mpc
comoving.  The redshift difference for the same radial separation is
$\Delta z=0.04$\citep{1999ApJS..120...49P}, which is a bandwidth of
500 khz.  Initial PAST will have an effective aperture of 40,000
m$^2(140 {\rm Mhz}/\nu)^2$, which corresponds to an aperture dilution
$\eta_A$ of about 1/100.  This results in a noise level after a year's
operation of $T_{sys}/(\eta_A\sqrt{t\Delta \nu}) \sim 10 $mK per
synthesized beam of three arc minutes.  Current models place the size
of ionization spheres at similar scales as this resolution, so one
expect of order unity signal to noise.  Radially, the resolution is
much better, so one could realistically expect 100 independent radial
modes in each synthesized beam across the epoch of reionization.  If
each radial mode represents a signal to noise of order unity, our
sensitivity is about $\Delta \kappa \sim 0.1$.  The square is the
lensing noise shown in Figure \ref{fig:kps} at the corresponding
angular scale $l\sim 3000$, and scales as white noise.

For gravitational lensing, one optimizes throughput by using the scale
where the signal-to-noise is of order unity.  Our procedure is to
square the density map at that scale, and use the changes in this
variance (square) to estimate the lensing strength $\kappa$.  The
power spectrum of $\kappa$ is now the two point correlation of our
density variance, which is related to a four point function of the
underlying density field.  Because of our large radial average of
squares, we assume that the contribution from the intrinsic spatial
connected four point function is neglibible.  With this procedure our
lensing noise power scales $\propto l^{2}$.

The actual sensitivity of the telescope to neutral hydrogen density
structures as a function of scale in the source plane is a different
calculation.  It depends on the actual layout of the elements, and the
number of elements at a given separation.  If we keep the collecting
area, and just shrink the baselines, the aperture dilution $\eta_A$
decreases as the square of the separation, while the radial scale is
proportionate to the angular scale.  The noise scales as
$\theta^{-2.5}$.  At double the angular scale, corresponding to six
arc minutes, this corresponds to a noise level of $13$ mK $\sqrt{{\rm
week}/t}$, so PAST can map out larger Stromgren spheres directly.  It
has been proposed that the ionization spheres up to $50$ Mpc have been
observed in existing QSO absorption line data
\citep{2004astro.ph..1188W}.  These would be observable at high signal
to noise with PAST.  LOFAR has a similar effective aperture, and we
would expect a similar sensitivity.  The second generation PAST (which
we call PAST+ in this paper) is expected to have ten times the
resolution and one hundred times the effective aperture.  This is four
times larger than the SKA, so for estimation purpose we clump these
two instruments into the same class, as shown by the more sensitive
noise lines in Figure \ref{fig:kps}.  To estimate the total
sensitivity on the full field power spectrum, we assumed that the
PAST array instantaneously measures 100 square degrees of the sky, and we
measure a 30\% redshift depth.  For PAST+/SKA we used 10 square degrees.

The problem can be broken into two parts: one must measure the power
spectrum of the 21cm emission, and then look for spatial variations in
the power.  Figure \ref{fig:mpower} shows the expected matter power
spectrum.  We show the noise of PAST/LOFAR as the dashed line. We also
plotted the statistical accuracy of the power spectrum measurement as
dotted lines. The matter power spectrum itself can of course also
provide cosmological information.  Unfortunately, this will again be
limited by the complication of gastrophysical processes such as
radiative transfer and star formation, which will modulate the spatial
distribution of neutral hydrogen.

\section{Processing Challenges}

Several problems arise in the endeavor to map gravitational lensing to
such an ambitious scale.   \citet{2003MNRAS.346..871O} have shown
that fluctuations in synchrotron emission from ionized gases can
outshine the fluctuations from the 21 cm emission.  As pointed out by
 \citet{2004MNRAS.347..187F}, this can probably be addressed by
removing all power-law spectrum spatial fluctuations.  A similar
requirement holds for the modelling of radio point sources.  These
sources are valuable to calibrate the point-spread function.  A
promising strategy is to use the brightest point sources in the map to
calibrate the spectrum.  The radio sources are emit synchrotron
radiation from relativistic electrons, which has no spectral
structure.  The task of looking for spectral features now consists of
comparing the spectrum of points on the map with that of nearby bright
sources.  This can be done to high accuracy.

Computationally, a brute force correlation of a very large number of
dipoles can be expensive. A direct correlation for a full sky
synthesis with $N$ dipoles with involves ${\cal O}(N^2)$ correlations.
Some designs allow beam forming within each synthesis node.  For nodes
with $n$ dipoles placed on regularly spaced patterns, the cost to form
the $n$ independent beams can be reduced to ${\cal O}(n\log n)$, so
the total correlation cost is ${\cal O}((N^2/n)\log n)$.  Measuring
the all-sky actual power spectra can also be computationally
challenging, especially if corrections to the weak lensing
approximation is desired \citep{2000ApJ...534L..19P}.  The general
theoretical framework is well understood \citep{Okamoto:2003zw}, but a
detailed computational pipeline will require significant software
development \citep{2003MNRAS.346..619P}. 

Since the array is sparse, not all projected baselines are measured
simultaneously.  Earth rotation fills in the map.  To perform
statistics on $10^{18}$ sources of course will require a
correspondingly large catalog.  The information is measured in the
exabytes.  With today's hard drives, this would cost billions of
dollars.  But storage and computation both follow Moore's law, and
will become affordable in the forseeable future.  Certainly many
challenges must be overcome to implement the ultimate array.

\section{Cosmological Constraints}

We discuss some of the cosmological constraints achievable with
pre-reionization 21cm emission.  The cosmic shear can be measured to
unity signal-to-noise for $l<500$ for PAST/LOFAR, $l<10000$ using
PAST+/SKA and $l<10^7$ using the ultimate detector.

Nominally, this results in an accuracy of projected dark matter power
spectrum determinations to better than $10^{-4}$ for PAST+/SKA,
and better than $10^{-7}$ in principle.  If the only uncertainty came
from the neutrino mass, we can apply equation (6) from
\citet{1999A&A...348...31C} and derive a nominal accuracy on the
neutrino mass of 0.1meV for PAST+/SKA and $\mu$ eV as the ultimate
limit.  This is already significantly more accurate than the mass
differences between neutrinos, and it should be possible to measure the
individual masses of each of the neutrino generations separately.
These masses can then be compared to experiment.

Over a fiducial range of source
redshifts $8<z<10$, the angular diameter distance varies by 5\%.  For
a lens half way between us and the source $z\sim 1.6$, the lensing
strength changes by 3\%, which is easily measurable.  Similarly,
tomography allows the reconstruction of the power spectrum at two
different redshifts.  For a lens screen 1/3 and 2/3 to the source
screen (redshifts 0.9 and 2.7 respectively), the relative amplitudes
are only 90\% degenerate, loosing only a factor of ten in accuracy for
determining the difference in amplitude.  The change in power spectrum
can be measured to an accuracy of $10^{-3}$, which allows very
accurate dynamical constraints on the equation of state of the
universe \citep{2002PhRvD..66h3515H}.  Most of the proposed
cosmological constraints by precision measurement of the dark matter
power spectrum can be implemented in this precision tomographic
survey.  Baryon oscillations have been suggested as an example of a
standard ruler \citep{2003ApJ...594..665B}.

Inflation generically predicts the existence of gravity waves produced
by Hawking radiation from the de Sitter horizon at the time of
inflation.  The amplitude at the quadrupole is $C_l \sim (H_I /M_{\rm
planck})^2$, where $H_I$ is the Hubble parameter during inflation.
Using $l^3\sim 10^{12}$ effectively independent sources in PAST+/SKA,
we can constrain $H_I <10^{-6} M_{\rm planck}$
\citep{2003PhRvL..91b1301D}.  The 21cm fluctuations should be at least
2 orders of magnitude more sensitive than the galaxy statistics used
in \citet{2003PhRvL..91b1301D}.  For the ultimate survey, we expect to
see $10^{18}$ minihalos on the sky.  If one resolves each, and
measures weak lensing, we can measure inflationary expansions all the
way to a Hubble parameter at inflation of $H_I\sim 10^{-9} M_{\rm
planck}$.

Such a precise measurement of the power spectrum would also unveal
second order effects.  Perturbations at inflation have amplitudes of
order ${\cal O} (10^{-5})$.  Various second order effects enter at an
amplitude of $10^{-10}$, and any measurement of the power spectrum
more accurate than one part in $10^5$ will depend on the second order
initial conditions.

These estimates are all idealized, and signficant work is needed both
in the instrument design and the processing stages to achieve these
sensitivities.

\section{Conclusions}

I have presented sensitivity estimates for epoch of reionization 21cm
weak lensing surveys.  We find that a precise mapping of the projected
dark matter structure on the sky can be achieved with several projects
under construction and planning, including PAST, its second generation
PAST+, LOFAR, and SKA.  In principle, it is also possible to image up
to $10^{18}$ minihalos with an extremely large dipole array.  Such a
future step would require an effective aperture of $\gtrsim 10^5$
km$^2$.  The overwhelming statistical accuracies would allow accurate
measurement or stringent upper bounds on primordial gravity waves, and
second order effects from the epoch of inflation.

Even with the upcoming experiments, exquisite maps of dark matter and
its evolution is possible, at much higher precision than any other
planned or proposed procedure.  The two dimensional dark matter
distribution can be measured to a signal-to-noise of better than unity
at scales down to $l\lesssim 10000$.  With the leverage of a redshift
range $8<z<10$, the evolution of the power spectrum, dark energy and
neutrino mass can be measured to better than percent accuracies.

I thank Martin White for helpful discussions and Pengjie Zhang for
the lensing code.

\newcommand{\apjl}{ApJ}      
\newcommand{\mnras}{MNRAS}   
\newcommand{\aap}{A\&A}

\bibliography{penbib}
\bibliographystyle{elsart-harv}

\appendix

\end{document}